\documentclass[12pt]{amsart}
\usepackage{amssymb}
\usepackage{color}
\usepackage{amsmath,epic,curves}
\usepackage[english]{babel}
\usepackage{graphicx}
\begin{document}
\baselineskip 6.0 truemm
\parindent 1.5 true pc

\newcommand\lan{\langle}
\newcommand\ran{\rangle}
\newcommand\tr{{\text{\rm Tr}}\,}
\newcommand\ot{\otimes}
\newcommand\ol{\overline}
\newcommand\join{\vee}
\newcommand\meet{\wedge}
\renewcommand\ker{{\text{\rm Ker}}\,}
\newcommand\image{{\text{\rm Im}}\,}
\newcommand\id{{\text{\rm id}}}
\newcommand\tp{{\text{\rm tp}}}
\newcommand\pr{\prime}
\newcommand\e{\epsilon}
\newcommand\la{\lambda}
\newcommand\inte{{\text{\rm int}}\,}
\newcommand\ttt{{\text{\rm t}}}
\newcommand\spa{{\text{\rm span}}\,}
\newcommand\conv{{\text{\rm conv}}\,}
\newcommand\rank{\ {\text{\rm rank of}}\ }
\newcommand\re{{\text{\rm Re}}\,}
\newcommand\ppt{\mathbb T}
\newcommand\rk{{\text{\rm rank}}\,}
\newcommand\bcolor{\color{blue}}
\newcommand\ecolor{\color{black}}
\newcommand\sss{\omega}

\title{Geometry for separable states and construction of entangled states with positive partial transposes}

\author{Kil-Chan Ha and Seung-Hyeok Kye}
\address{Faculty of Mathematics and Applied Statistics, Sejong University, Seoul 143-747, Korea}
\address{Department of Mathematics and Institute of Mathematics\\Seoul National University\\Seoul 151-742, Korea}

\date\today

\thanks{partially supported by NRFK 2013-020897 and NRFK 2013-004942}



\begin{abstract}
We construct faces of the convex set of all $2\otimes 4$ bipartite
separable states, which are affinely isomorphic to the simplex
$\Delta_{9}$ with ten extreme points. Every interior point of these
faces is a separable state which has a unique decomposition into
$10$ product states, even though ranks of the state and its partial
transpose are $5$ and $7$, respectively. We also note that the
number $10$ is greater than $2\times 4$, to disprove a conjecture on
the lengths of qubit-qudit separable states. This face is inscribed
in the corresponding face of the convex set of all PPT states so that
sub-simplices $\Delta_k$ of $\Delta_{9}$ share the boundary if and
only if $k\le 5$. This enables us to find a large class of $2\otimes
4$ PPT entangled edge states with rank five.
\end{abstract}

\maketitle

\section{Introduction}\label{sec:intro}
One of the fundamental question in the theory of quantum entanglement is how to
distinguish and construct entangled states. Even though the
PPT criterion \cite{choi-ppt,peres} gives us a simple and powerful
necessary condition for separability together with the range
criterion \cite{p-horo}, it is not clear how to distinguish
entanglement satisfying these two criteria. One of the possible way
to overcome this difficulty is to compare the geometries for
separable states and PPT states, as it was suggested in a recent
work \cite{ha_kye_uni_decom}.

We note that the set of all separable states (respectively PPT
states) makes a convex set, which will be denoted by $\mathbb S$
(respectively $\mathbb T$). In order to  understand the geometry of
a convex set, we need to characterize the facial structures. The
facial structures for the convex set $\mathbb T$ is relatively well
understood \cite{ha_kye_04}. It is also known \cite{choi_kye} that a
given PPT state $\varrho$ satisfies the range criterion if and only
if the face of $\mathbb T$ determined by $\varrho$ has a separable
state in its interior. If we understand the facial structures of the
corresponding face of $\mathbb S$, then it is easy to distinguish
and construct entangled states within the face of $\mathbb T$. This
is the case when the corresponding face for $\mathbb S$ is affinely
isomorphic to a simplex.

The authors \cite{ha_kye_uni_decom} exploited this idea for the $3\otimes 3$ case, to construct
faces of $\mathbb S$ which are isomorphic to the simplex $\Delta_5$ with six extreme points,
and understand how PPT entangled edge states of rank four arise. This construction also gives
examples of separable states whose lengths are greater than the maximum of ranks of themselves and
their partial transposes. The main idea was to begin with generic $5$-dimensional
subspaces of $\mathbb C^3\ot \mathbb C^3$ which has six product vectors, and exploit the fact that
the number of product vectors is greater than the dimension.

In this paper, we pursue the same idea for the $2\ot 4$ case, which is the smallest dimension
where PPT entangled states arise. But, the above idea for the $3\ot 3$ case does not work for this case, because
the number $d$ is the dimension for generic subspaces of $\mathbb C^2\ot\mathbb C^d$ with
finitely many product vectors, and generic $d$-dimensional subspaces have just $d$ product vectors.
To overcome this difficulty, we consider the equation
\begin{equation}\label{equation}
|x\ot y\rangle\ \in D,\qquad |\bar x\ot y\rangle\ \in E
\end{equation}
for a given pair $(D,E)$ of subspaces of $\mathbb C^2\ot\mathbb C^4$, where
$|x\otimes y\rangle:=|x\rangle \otimes |y\rangle$ and $|\bar x\rangle$ denotes the conjugate of $|x\rangle$.
We construct $5$-dimensional spaces $D$ and $7$-dimensional spaces $E$, for which the above equations
have exactly $10$ solutions. This enables us to construct faces of $\mathbb S$ isomorphic to the
simplex $\Delta_{9}$ with ten extreme points.

Any interior point of the face $\Delta_{9}$ is a separable state
with the unique decomposition into ten product states, and has the
length ten. This disproves the conjecture \cite{chen_dj_semialg}
which claims that lengths of $2\ot d$ separable states are at most
$2\times d$. We recall that a PPT state $\varrho$ is of type $(p,q)$
if the ranks of $\varrho$ and $\varrho^\Gamma$ are $p$ and $q$,
respectively. We also note that the boundary of this face $\Delta_9$
consists of simplices $\Delta_{k}$ with $k+1$ extreme points, for
$k\le 8$. By the construction, every interior point of the face
$\Delta_{9}$ is a separable state of type $(5,7)$. We show that any
choice of seven product vectors $|\bar x\ot y\rangle$ among ten
solutions are linearly independent. From this, we conclude that if a
boundary point $\varrho_1$ of $\Delta_{9}$ is in the interior of
$\Delta_{k}$ with $6\le k\le 8$ then the line segment from an interior point
$\varrho_0$ of $\Delta_9$ to $\varrho_1$ can be extended within the convex set
$\mathbb T$, to get PPT entangled states of rank five. For known
examples of $2\ot 4$ PPT entangled states, see \cite{agkl,p-horo}.

In the next section, we briefly review the material behind the above idea we have just explained, and give the construction
in the Section 3.

\section{Background}

A density matrix $\varrho$ in the tensor product $M_m\ot M_n$
of matrix algebras is said to be separable if it is the convex combination of
product states, and so it is of the form
\begin{equation}\label{sep}
\varrho=\sum_{i=1}^k \lambda_i |x_i\ot y_i\rangle\langle x_i\ot y_i|,
\end{equation}
with unit product
vectors $|x_i\ot y_i\rangle$ in the space $\mathbb C^m\ot \mathbb C^n$ and
positive numbers $\lambda_i$ with $\sum_{i=1}^k\lambda_i=1$.
A non-separable state is called entangled. Because the partial transpose $\varrho^\Gamma$ of the state $\varrho$
in (\ref{sep}) is given by
$$
\varrho^\Gamma=\sum_{i=1}^k \lambda_i |\bar x_i\ot y_i\rangle\langle \bar x_i\ot y_i|,
$$
we see that the partial transpose of a separable state is also positive. This is the PPT criterion \cite{choi-ppt,peres}.
Furthermore, we also see that if $\varrho$ is separable then there must exist product vectors
$|x_i\ot y_i\rangle$ satisfying
\begin{equation}\label{range}
{\mathcal R}\varrho=\spa\{|x_i\ot y_i\rangle\},\qquad
{\mathcal R}\varrho^\Gamma=\spa\{|\bar x_i\ot y_i\rangle\},
\end{equation}
as the range criterion \cite{p-horo} tells, where ${\mathcal R}\varrho$ denotes the range space of the state $\varrho$.

A convex subset $F$ of a convex set $C$ is said to be a face if
it satisfies the condition:
If a point in $F$ is a convex combination of two points in $C$ then they must be points of $F$.
A face consisting of a single point is call an extreme point. A point $x$ in a convex set $C$ is said to be
an interior point of $C$ if it is an interior point of $C$ with respect to the relative topology
of the affine manifold generated by $C$. It is well known that every convex set is completely partitioned
into interiors of faces. In this sense, a point $x$ in a convex set determines a unique face in which $x$ is an interior
point. This is the smallest face containing $x$. A point of $C$ is called a boundary point if it is not an interior point.

Any face of $\mathbb T$ is determined \cite{ha_kye_04} by a pair $(D,E)$ of subspaces of $\mathbb C^m\ot\mathbb C^n$, and is of the form
$$
\tau(D,E)=\{\varrho\in\mathbb T: {\mathcal R}\varrho\subset D,\ {\mathcal R}\varrho^\Gamma \subset E\}.
$$
Conversely, the set $\tau(D,E)$ is a face unless it is empty.
The interior of $\tau(D,E)$ is given by
$$
\inte\tau(D,E)=\{\varrho\in\mathbb T: {\mathcal R}\varrho= D,\ {\mathcal R}\varrho^\Gamma = E\}.
$$
It was also shown in \cite{choi_kye} that a PPT state $\varrho$ satisfies the range criterion if and only if
the interior of the face $\tau(D,E)$ of $\mathbb T$ determined by $\varrho$ has a separable state. In this case, the
face $\mathbb S\cap\tau(D,E)$ of $\mathbb S$ share interior points with the face $\tau(D,E)$ of $\mathbb T$.
Therefore, it is crucial to understand the facial structures of $\mathbb S\cap\tau(D,E)$ in order to
determine if $\varrho$ is separable or not.

The study of facial structures of $\mathbb S$ has been initiated by Alfsen and Schultz \cite{alfsen},
where they searched for faces of $\mathbb S$ which is affinely isomorphic to a simplex. Suppose that
a convex set $C$ is on the hyperplane of codimension one in the $(d+1)$-dimensional real vector space which does not
contain the origin. Then $C$ is a simplex if and only if it is the convex hull of $d+1$ linearly independent points on the hyperplane.
This simplex will be denoted by $\Delta_d$. Therefore, if a separable state in (\ref{sep}) determines a face,
then it is isomorphic to a simplex if and only if the product states in the expression are linearly independent in the
real vector space of Hermitian matrices. For further progress on the facial structures for separable states, see
\cite{alfsen_2,choi_kye,ha_kye_uni_decom,hansen_ext_wit,kye_trigono}.
The length of a separable state $\varrho$ is defined by the smallest number $k$ with which the expression (\ref{sep}) is possible.
It is clear that if a separable state determines the face isomorphic to the simplex $\Delta_k$ then it has the length $k+1$.

Now, we are ready to explain the main idea of the construction in the next section.
We construct a $5$-dimensional space $D$ and a $7$-dimensional space $E$
of $\mathbb C^2\ot\mathbb C^4$, and show the following:
\begin{enumerate}
\item[(i)]
The equation (\ref{equation}) has exactly ten solutions.
\item[(ii)]
The corresponding ten product states are linearly independent.
\item[(iii)]
Any choice of five product vectors $|x\ot y\rangle$ span the space $D$.
\item[(iv)]
Any choice of seven product vectors $|\bar x\ot y\rangle$ span the space $E$.
\end{enumerate}
We conclude that the face $\tau(D,E)$ has a separable state in the interior by (iii) and (iv), and the face
$\tau(D,E)\cap\mathbb S$ is affinely isomorphic to the simplex $\Delta_{9}$ by (i) and (ii).

We take an interior point $\varrho_0$ in the face $\tau(D,E)\cap\mathbb S$, which will be denoted by just $\Delta_{9}$,
and take a boundary point $\varrho_1$ which determines the face isomorphic to $\Delta_k$ with $k\le 8$. This means that
$\varrho_1$ is the convex combination of $k+1$ product states.
Consider $\varrho_t=(1-t)\varrho_0+t\varrho_1$ for $t>1$. If $k+1\le 6$ then ${\mathcal R}\varrho_1^\Gamma$ is a proper
subspace of ${\mathcal R}\varrho_0^\Gamma$, and so we see that $\varrho_t$ is never positive for $t>1$. If $k+1\ge 7$ then we see that
the range spaces of $\varrho_0$ and $\varrho_1$ coincide by (iii), and same for the range spaces of
$\varrho_0^\Gamma$ and $\varrho_1^\Gamma$ by (iv).
Therefore, we see that there exist $t>1$ such that $\varrho_t$ is of PPT. It is clear that this is an entangled state.
If we take the largest $t$ such that $\varrho_t$ is of PPT then
$\varrho_t$ is of type $(p,q)$ with $p<5$ or $q<7$. But, it is not possible to have $p<5$ by \cite{2xn},
because $\varrho_t$ is an entangled state. Therefore, we conclude that $\varrho_t$ is of type $(5,5)$ or $(5,6)$.

\section{Construction}

Let $D$ be the $5$-dimensional subspace of $\mathbb C^2\ot\mathbb C^4$ which is orthogonal to the following three
vectors:
$$
\begin{aligned}
|v_1\rangle &=(0,1,0,0,-1,0,0,0)^{\rm t},\\
|v_2\rangle &=(0,0,1,0,0,-1,0,0)^{\rm t},\\
|v_3\rangle &=(0,0,0,1,0,0,-1,0)^{\rm t}.
\end{aligned}
$$
We note that these three vectors span a completely entangled space which has no product vectors.
It is easy to see that every product vector $|z\rangle =|x\rangle \otimes |y\rangle$ in the space $D$ is one of the following forms:
\begin{equation}\label{eq:prod_vec}
|z_1\rangle =(0,1)^{\rm t}\otimes (0,0,0,1)^{\rm t},\quad |z(\alpha)\rangle =(1,\alpha)^{\rm t}\otimes (1,\alpha,\alpha^2,\alpha^3)^{\rm t}
\end{equation}
for a complex number $\alpha$. Note that the partial conjugate of $|z(\alpha)\rangle$, which will be denoted by
$|\bar z(\alpha)\rangle$, is given by
$$
|\bar z(\alpha)\rangle
=(1,\bar\alpha)^{\rm t}\otimes (1,\alpha,\alpha^2,\alpha^3)^{\rm t}
=(1,\alpha,\alpha^2,\alpha^3,\bar\alpha, |\alpha|^2, |\alpha|^2\alpha, |\alpha|^2\alpha^2)^{\rm t}.
$$

For given real numbers $a$ and $b$ with the relation $0<b<4a^3/27$, we consider the vector
\begin{equation}\label{vecw}
|w\rangle =(b,0,0,1,0,-a,0,0)^{\rm t}, 
\end{equation}
and let $E$ be the $7$-dimensional subspace of $\mathbb C^2\ot\mathbb C^4$ orthogonal to the vector $|w\rangle$.

Now, we proceed to solve the equation (\ref{equation}) for the above $D$ and $E$.
We note that the partial conjugate of $|z_1\rangle$ belongs to $E$, and so $|z_1\rangle$ is a solution.
In order to find complex numbers $\alpha$ so that $|z(\alpha)\rangle$ is a solution,
we solve the equation $\langle \bar z(\alpha)|w\rangle=0$, that is,
\begin{equation}\label{eq:complex}
b+\alpha^3-a |\alpha|^2=0.
\end{equation}
We first note that $\alpha^3$ must be a real number, and so we have $\alpha=r e^{i\theta}$ with $3\theta=n\pi$ and $r>0$.
If $n$ is an even integer, then the equation \eqref{eq:complex} is reduced to
\[
r^3-a r^2+b=0,
\] and we get two distinct positive roots $r_1$ and $r_2$ from the condition $0<b<4a^3/27$. In the case of an odd integer $n$,
we get one positive root $r_3$ of the equation $r^3+a r^2-b=0$ by the same condition.
We also note that $r_1,r_2,r_3$ are mutually distinct.
Therefore, we have the following nine solutions of the equation \eqref{eq:complex}:
\begin{equation}\label{eq:sols}
r_1,\ r_1 \omega,\ r_1\omega^2,\ r_2,\ r_2\omega,\ r_2\omega^2,\ -r_3,\ -r_3 \omega,\ -r_3\omega^2,
\end{equation}
where $\omega$ is the third root of unity.
For the notational convenience, we rewrite the normalizations of the product vectors $|z(\alpha)\rangle$
in \eqref{eq:prod_vec} for the above nine $\alpha$'s
by $|z(\alpha_i)\rangle$ for $i=2,3,\cdots,10$.

In order to show the item (ii) in the last section, suppose that
\[
a_1 |z_1\rangle \langle z_1|+\sum_{i=2}^{10} a_i |z(\alpha_i)\rangle \langle z(\alpha_i)|=O,
\]
where $O$ is the $8\times 8$ zero matrix.
Note that $|z(\alpha)\rangle \langle z(\alpha)|$ is given by
$$
|z(\alpha)\rangle \langle z(\alpha)|
=\begin{pmatrix}
1 & \bar\alpha & \bar\alpha^2 & \bar\alpha^3 & \bar\alpha & \bar\alpha^2 & \bar\alpha^3 &\bar\alpha^4\\
\alpha & |\alpha|^2 & && \cdots &&& |\alpha|^2\bar\alpha^3\\
\vdots &\vdots     & && \ddots &&&\vdots\\
\alpha^4 & |\alpha|^2\alpha^3 & && \cdots &&& |\alpha|^8
\end{pmatrix},
$$
which is an $8\times 8$ matrix.
Therefore, we get sixty four linear equations with respect to $a_i \ (i=1,2,\cdots, 10)$
by comparing the entries of the both sides.
If we write
$$
C =
\begin{pmatrix}
0&1 & 1 & \cdots & 1 & 1 \\
0&\bar\alpha_2 & \bar\alpha_3  & \cdots & \bar\alpha_9 & \bar\alpha_{10}  \\
\vdots &\vdots & \vdots & \ddots  & \vdots & \vdots   \\
0&|\alpha_2|^6 & |\alpha_3|^6 & \cdots & |\alpha_9|^6 &  |\alpha_{10}|^6 \\
\vdots &\vdots & \vdots & \ddots  & \vdots & \vdots   \\
1&|\alpha_2|^8 & |\alpha_3|^8 & \cdots & |\alpha_9|^8 &  |\alpha_{10}|^8
\end{pmatrix}
$$
which is a $64\times 10$ matrix, and $A=(a_1,a_2,\cdots,a_{10})^{\rm t}$, then we have the equation
$$
CA=O.
$$
We note that entries of the first  column of $C$ are all zero except the last entry.
Since we have
$$
|\alpha_i|^8=a^2 |\alpha_i|^6 -2ab|\alpha_i|^4+b^2|\alpha_i|^2,\qquad i=2,3,\dots,10
$$
from the equation
\eqref{eq:complex}, we can conclude that the row vector
$(0,|\alpha_2|^8,|\alpha_3|^8,\cdots,|\alpha_{10}|^8)$ is
the linear combination of three rows of the above matrix $C$. Therefore, we have $a_{1}=0$.
Since any nine product states corresponding to the nine product vectors
are linearly independent by Proposition 2.2 in \cite{ha_kye_uni_decom}, we have $a_2=a_3=\cdots =a_{10}=0$.
Proposition 2.1 of \cite{ha_kye_uni_decom} also tells us that any choice of five product vectors among ten solutions
are linearly independent.

It remains to show the item (iv) in the last section.
Without loss of generality, it suffices to consider the following five cases:
\begin{itemize}
\item[(i)] $\{ |z_1\rangle,|z(r_1)\rangle,|z(r_1 \omega)\rangle,|z(r_1\omega^2)\rangle,|z(r_2)\rangle,|z(r_2\omega)\rangle,|z(r_2\omega^2)\rangle\}$,
\item[(ii)] $\{ |z_1\rangle,|z(r_1)\rangle,|z(r_1 \omega)\rangle,|z(r_1\omega^2)\rangle,|z(r_2)\rangle,|z(r_2\omega)\rangle,|z(-r_3)\rangle\}$,
\item[(iii)] $\{ |z_1\rangle,|z(r_1)\rangle,|z(r_1 \omega)\rangle,|z(r_2)\rangle,|z(r_2\omega)\rangle,|z(-r_3)\rangle,|z(-r_3\omega)\rangle\}$,
\item[(iv)] $\{|z(r_1)\rangle,|z(r_1 \omega)\rangle,|z(r_1\omega^2)\rangle,|z(r_2)\rangle,|z(r_2\omega)\rangle,|z(r_2\omega^2)\rangle,|z(r_3)\rangle\}$,
\item[(v)]$\{ |z(r_1)\rangle,|z(r_1 \omega)\rangle,|z(r_1\omega^2)\rangle,|z(r_2)\rangle,|z(r_2\omega)\rangle,|z(r_3)\rangle,|z(r_3\omega)\rangle\}$,
\end{itemize}
For each case, we form the $7\times 8$ matrix whose rows are given by the seven product vectors identified with row vectors in $\mathbb C^8$.
Then, in any cases, it is easy to see that the reduced row echelon form of the matrix is given by
\[
\begin{pmatrix}
1 & 0 & 0 & 0 & 0 & b/a & 0 & 0\\
0 & 1 & 0 & 0 & 0 & 0 & 0 & 0\\
0 & 0 & 1 & 0 & 0 & 0 & 0 & 0 \\
0 & 0 & 0 & 1 & 0 & 1/a & 0 & 0\\
0 & 0 & 0 & 0 & 1 & 0 & 0 & 0\\
0 & 0 & 0 & 0 & 0 & 0 & 1 & 0\\
0 & 0 & 0 & 0 & 0 & 0 & 0 & 1
\end{pmatrix}.
\]
Therefore, for any choice of seven product vectors among ten solutions, we see that
the corresponding seven partial conjugates are linearly independent, and so they span the space $E$.

Finally, we illustrate the above discussion with an explicit example.
We consider the case of $a=2$ and $b=1$ in \eqref{vecw}. From the equation $r^3-2r^2+1=0$,
we get two positive solutions $r_1=1$ and $r_2=(1+\sqrt 5)/2$.
We also get one positive solution $r_3=(\sqrt{5}-1)/2$ from the equation $r^3+2r^2-1=0$.

By a direct computation, we have an interior point $\varrho_0$ of $\Delta_9$:
\[
\begin{aligned}
\varrho_0
&=\frac 1{10} \left(|z_1\rangle \langle z_1|+\sum_{i=2}^{10} |z(\alpha_i)\rangle \langle z(\alpha_i)|\right)\\
&=
\frac 1{400}\begin{pmatrix}
71 & 0 & 0 & 7 & 0 & 0 & 7 & 0\\
0 & 39 & 0 & 0 & 39 & 0 & 0 & 23\\
0 & 0 & 31 & 0 & 0 & 31 & 0 & 0 \\
7 & 0 & 0 & 39 & 0 & 0 & 39 & 0\\
0 & 39 & 0 & 0 & 39 & 0 & 0 & 23\\
0 & 0 & 31 & 0 & 0 & 31 & 0 & 0 \\
7 & 0 & 0 & 39 & 0 & 0 & 39 & 0\\
0 & 23 & 0 & 0 & 23 & 0 & 0 & 111
\end{pmatrix},
\end{aligned}
\]
where $|z(\alpha_i)\rangle$ is the normalized product vectors with $\alpha_i$'s in \eqref{eq:sols}.
We consider the $8$-simplex $\Delta_8$ determined by these nine product vectors $|z(\alpha_i)\rangle$'s.
An interior point $\varrho_1$ of this face is given by
\[
\varrho_1
=\frac 1 9\sum_{i=2}^{10}  |z(\alpha_i)\rangle \langle z(\alpha_i)|
=
\frac 1{360}
\begin{pmatrix}
71 & 0 & 0 & 7 & 0 & 0 & 7 & 0\\
0 & 13 & 0 & 0 & 13 & 0 & 0 & 23\\
0 & 0 & 31 & 0 & 0 & 31 & 0 & 0 \\
7 & 0 & 0 & 13 & 0 & 0 & 13 & 0\\
0 & 13 & 0 & 0 & 13 & 0 & 0 & 23\\
0 & 0 & 31 & 0 & 0 & 31 & 0 & 0 \\
7 & 0 & 0 & 13 & 0 & 0 & 13 & 0\\
0 & 23 & 0 & 0 & 23 & 0 & 0 & 71
\end{pmatrix}.
\]
We put $\rho_t=(1-t)\varrho_0+t\varrho_1$, and explore eigenvalues of $\rho_t$ and $\rho_t^{\Gamma}$.
We denote by $\lambda_i(t)$ and  $\mu_i(t)$  the eigenvalues of $\rho_t$ and   $\rho_t^{\Gamma}$, respectively.
Then, we see that there exist $\nu\approx 1.48192$ so that $\rho_{\nu}$ is on the boundary of the face $\tau(D,E)$,
which is a PPT entangled edge state of type $(5,6)$.
We note that $\rho_t$ is a  PPT entangled state of type $(5,7)$ for $1<t<\nu$.
\begin{figure}[h!]
\begin{center}
\includegraphics[scale=0.4]{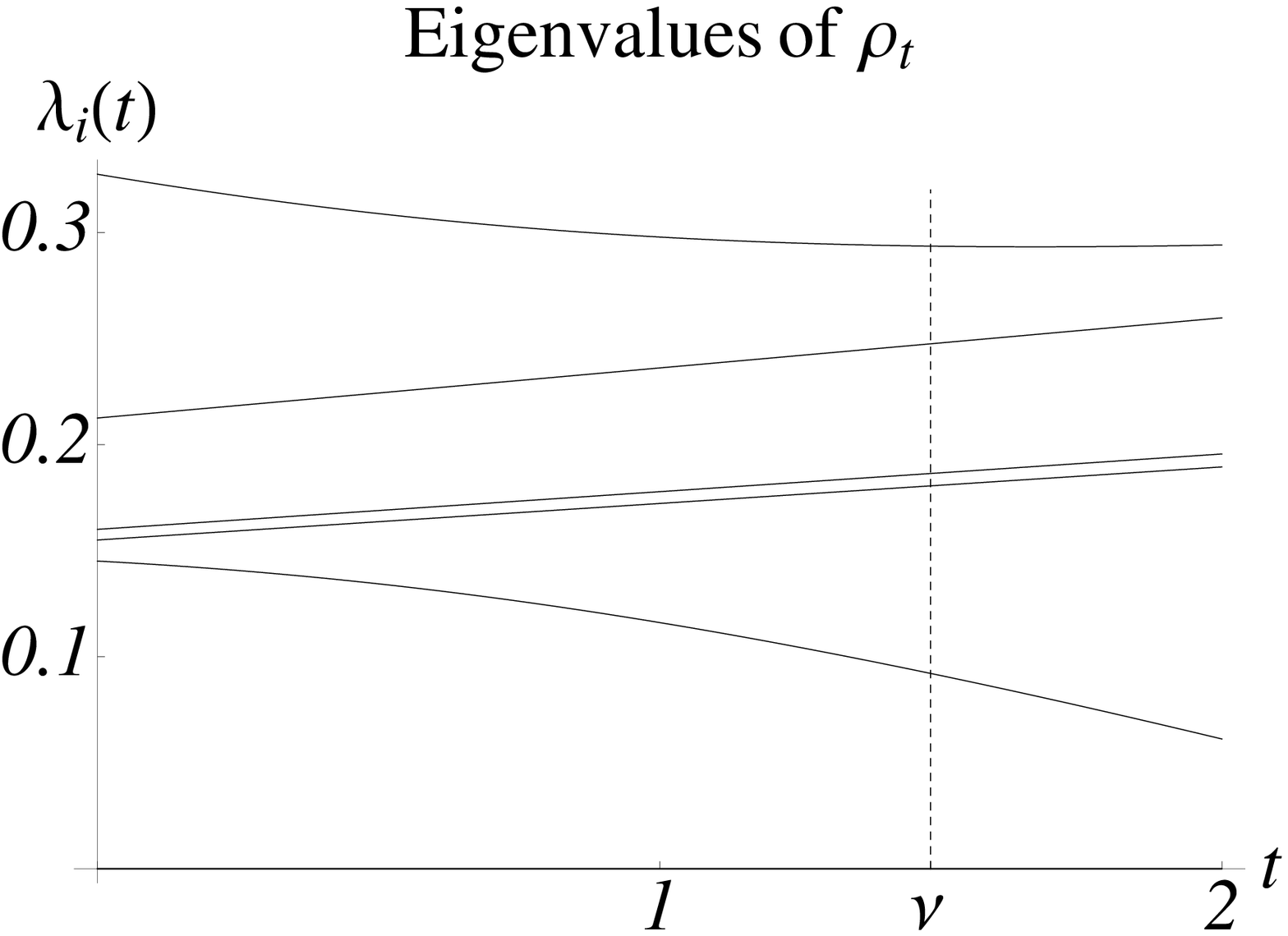}
\includegraphics[scale=0.4]{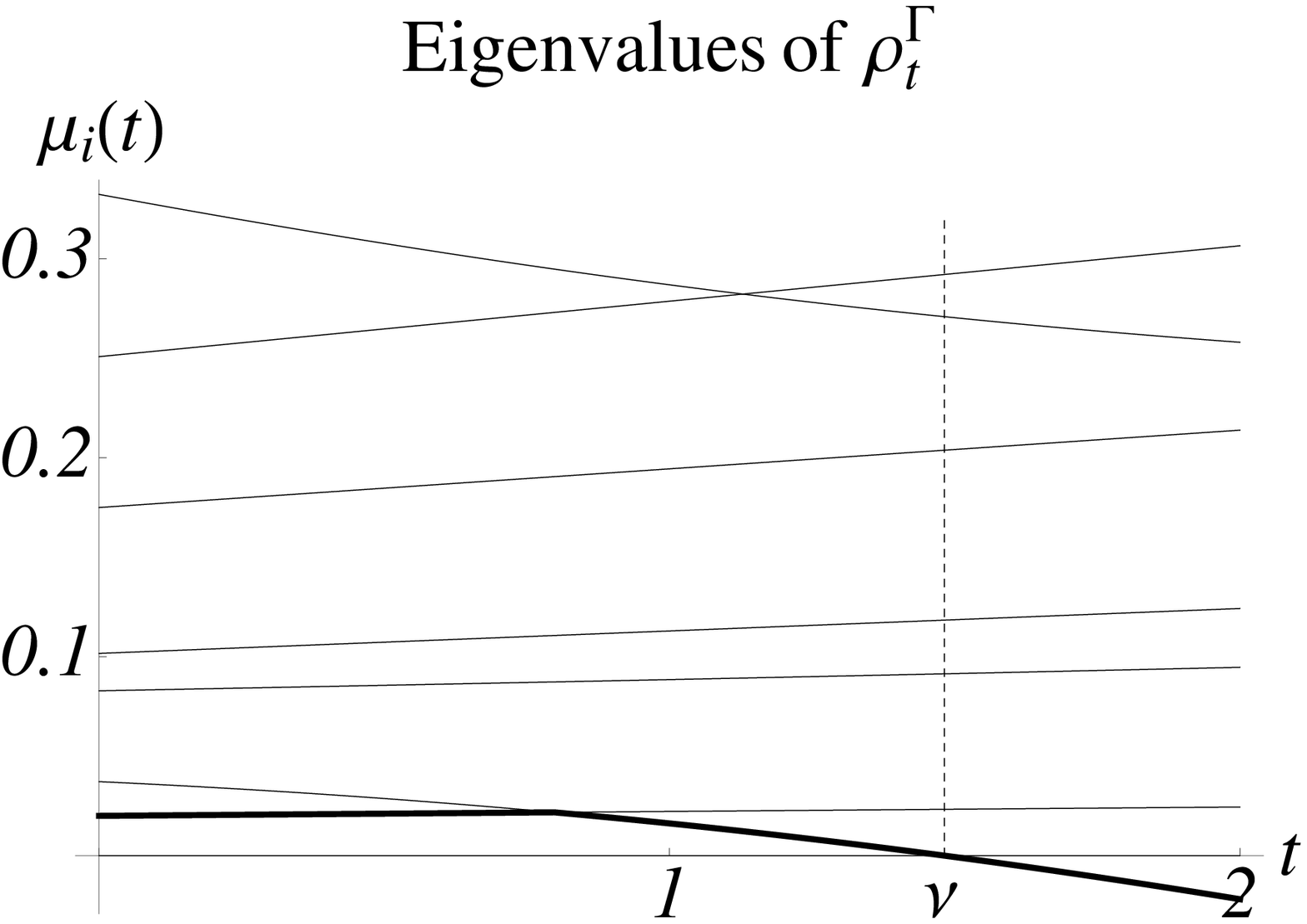}
\end{center}
\caption{In both graphs, each curve represents distinct eigenvalues. So, the rank of $\rho_{\nu}$ is
five and the rank of $\rho_{\nu}^{\Gamma}$ is six. In the graphs of $\mu_i$, the thick line
denotes the smallest eigenvalue of $\rho^{\Gamma}$ except zero.  }
\end{figure}

\section{Conclusion}\

In this paper, we have constructed faces of the convex set of all $2\ot 4$ separable states, which are isomorphic to
the simplex $\Delta_9$.
The boundary of this face consists of simplices $\Delta_k$ with $k\le 8$. Note that the number of faces isomorphic to $\Delta_k$ is
$\binom {10}{k+1}$. The discussion in Section 2 tells us that the interior of $\Delta_k$ is located in the interior of the face $\tau(D,E)$
if and only if $k\ge 6$. If $k\le 5$ then $\Delta_k$ is located on the boundary of $\tau(D,E)$. Since every interior point of
$\Delta_5$ is a separable state of type $(5,6)$, it is very plausible that the boundary point $\varrho_t$ of $\tau(D,E)$ is
also of type $(5,6)$. Actually, we got a PPT entangled state of type $(5,6)$ in the last numerical examples.
It is clear that the PPT entangled states which are located on the boundary of the face $\tau(D,E)$ must be edge states,
but it is not clear if they are extreme or not.
It would be interesting if we may get PPT entangled states of type $(5,5)$ by a similar construction.

\end{document}